# Vortex-like surface wave and its role in the transient phenomena of meta-material focusing

Lei Zhou [a,b]   and C. T. Chan [a]

[a] *Physics Department, Hong Kong University of Science and Technology, Clear Water Bay, Kowloon, Hong Kong, China*

[b] *Surface Key Laboratory and Physics Department, Fudan University, Shanghai 200433, P. R. China*

We show that a slab of meta-material (with $\varepsilon = \mu = -1 + i\Delta$) possesses a vortex-like surface wave with no ability to transport energy, whose nature is completely different from a localized mode or a standing wave. Through computations based on a rigorous time-dependent Green's function approach, we demonstrate that such a mode inevitably generates characteristic image oscillations in two-dimensional focusing with even a *monochromatic* source, which were observed in many numerical simulations, but such oscillations are weak in three-dimensional focusing.

Veselago proposed that a flat slab of material with $\varepsilon = \mu = -1$ could function as a lens to focus electromagnetic (EM) waves [1]. Pendry showed that the proposed lens was in fact a *perfect* one [2]. It was then argued that a small deviation in material properties (i.e. $\mu = -1 + \delta$ and/or $\varepsilon = -1 + \delta$ ) results in a *super* lens with imaging resolution beyond the usual diffraction limit [3,4]. Several studies, including both finite-difference-time-domain (FDTD) simulations [5-10] and theoretical analyses [4,11-14], have been performed on such *perfect* or *super* lenses. The results obtained so far are, however, not entirely consistent. While some studies confirmed the existences of perfect or super imaging [4,6-8,11], issues were raised by other authors, who showed that a perfect image could never be realized in practice [5,12] and no enhanced resolution was found in super-lens focusing [9].

We note that most studies to-date were interested in the stabilized image pattern under a given frequency except Ref. [14] where the relaxation was studied by an approximate model. Although transient effects [15] are less important for a conventional lens, we show they are crucial in super lens focusing and the super-resolution is intimately related to the transient phenomena. A pioneering FDTD simulation showed that no steady foci were found in perfect and super lens focusing and the fields varied dramatically over time [5]. A later FDTD work identified such dramatic field change as a characteristic oscillation [7]. Other FDTD researchers also observed such oscillations, and had to add absorptions to obtain stable images in their simulations [8,9]. Since the employed sources in all FDTD simulations are strictly *monochromatic* [5-9], the nature of this oscillation, with such a characteristic frequency, is rather intriguing. In this letter, we show that a slab of $\varepsilon = \mu = -1$ material possesses a *unique* vortex-like surface wave with no ability to transport energy. The nature of this state is different from either a localized mode or a standing wave. Through rigorous Green's function (GF) computations, we demonstrate that existence of such a mode inevitably leads to strong field oscillations in two dimensional (2D) focusing with even a *monochromatic* source, but such oscillations are weak in three dimensional (3D) focusing. We thus provide a natural explanation for many numerical simulations [5-9], and predict a pronounced effect of dimensionality dependence in the time evolution of optical effects in meta-material, and in particular, the super lens focusing.

We start from establishing a rigorous GF approach to study quantitatively the focusing effect. Using a dyadic Green's function $\overset{\leftrightarrow}{G}(r,r';t,t')$, the **E** field can be calculated as $\vec{E}(\vec{r},t) = -\mu \int \overset{\leftrightarrow}{G}(\vec{r},t;\vec{r}',t') \bullet \dot{\vec{J}}(\vec{r}',t') d\vec{r}' dt'$, where $\vec{J}(\vec{r},t)$ is a current source and the dot means time derivative. In the 2D case, we assume a line source that is located at the origin and operates from $t=0$, with a form $\vec{J}(\vec{r},t) = \hat{y} I_0 \delta(x)\delta(z) e^{-i\omega_0 t}\theta(t)$. A slab of meta-material of thickness $d$ is placed at the $xy$-plane between $z = -d/2$ and $z = -3d/2$ as a lens to focus EM waves radiated from the source into an image plane at $z = -2d$. The time-dependent **E** field can be written as

$$\vec{E}(\vec{r},t) = \frac{1}{2\pi} \int d\omega e^{-i\omega t} \vec{E}(\vec{r},\omega) \frac{\omega}{\omega - \omega_0 + i\eta} \quad , \qquad (1)$$

where $\vec{E}(\vec{r},\omega) = -\mu I_0 \int \overset{\leftrightarrow}{G}(\vec{r},y';\omega) \bullet \hat{y} \, dy'$ with $\overset{\leftrightarrow}{G}(\vec{r},\vec{r}';\omega)$ being the Fourier transform of $\overset{\leftrightarrow}{G}(\vec{r},t;\vec{r}',t')$, and $\eta$ is a positive infinitesimal number to ensure the causality. Following Ref. [13], we expand $\overset{\leftrightarrow}{G}(\vec{r},\vec{r}';\omega)$ in different regions with respect to parallel $\vec{k}$ components and polarizations, and obtain all components of $\vec{E}(\vec{r},\omega)$ in every region by matching boundary conditions. For example, on the $xz$-plane containing the source (i.e. $y=0$), we find for $z<$-$3d/2$ (i.e. the image region):

$$E_{2y}(x,z;\omega) = -\frac{i\mu_0 I_0}{4\pi} \int \frac{e^{ik_x x}}{k_{0z}} T^{TE} e^{-ik_{0z} z} dk_x \quad (2)$$

Here $k = \omega/c$, $k_{0z}^2 + k_\parallel^2 = (\omega/c)^2$, $T^{TE}$ is the transmission coefficient for incident EM waves with $\vec{k} = (\vec{k}_\parallel, k_z)$ and transverse-electric (TE) polarization (i.e. **E** parallel to the interface). We also investigate the 3D case with a point source of the form $\vec{J}(\vec{r}\,',t') = P_0 \hat{y} \delta(\vec{r}\,')e^{-i\omega_0 t'}\theta(t')$ [13]. Putting the frequency-dependent fields as Eq. (2) back into Eq. (1), we obtain time-dependent field values. It has been shown that $T^{TE}$ diverges at some specific $k_\parallel$ points, corresponding to the surface wave (SW) excitations of the slab [4]. With finite absorption, $|T^{TE}|$ is finite but still very large. The integration has to be done numerically, and convergence is assured by using an adaptive $k_\parallel$ mesh, with $\Delta k_\parallel \propto \left[1 + |T^{TE}(k_\parallel)|\right]^{-1}$, sampling more $k$ points around the singularity. A large maximum $k_\parallel$ value ($k_{max}$) is used, and the convergence against $k_{max}$ has been carefully checked. The same technique applies to the singularity encountered in the frequency integration (1).

We first consider the 2D case. In order to compare with previous studies [5-10], here we also consider a postulated artificial material with $\varepsilon = \mu = 1 - \dfrac{f_p^2}{f(f + i\gamma)}$ with $f_p = 10\sqrt{2}$ GHz. The working frequency is 10GHz at which $\varepsilon = \mu \simeq -1 + i\gamma/5$ when $\gamma$ is very small. For a slab with $d = 10$ mm and $\gamma = 0.005$ GHz, we plotted the calculated image resolution $w(t)$, defined as the peak width measured at half-maximum, and field amplitude $|E_y^{2D}|$ (at the image point) in Fig. 1(a) and (b). Results with propagating components only ($k_{max} = \omega/c$) are shown together for comparison. We find that the time evolution is dominated by damped oscillations with a characteristic period, confirming recent FDTD simulation results [7]. A thinner lens yields a higher-frequency oscillation [see the dashed line in Fig. 1(b) for $d = 8$ mm], and a larger $\gamma$ suppresses the oscillations better (see the dotted line for $\gamma = 0.02$ GHz). The oscillation is apparently contributed by the evanescent waves since results including only propagating waves do not show any oscillations [see circles in Fig. (1)].

The oscillation is closely related to the SW spectrum [16]. Figure 2(a) depicts the SW with TE polarization for two lenses with different thicknesses. The SW dispersion for a single air/lens interface is a straight parallel line at 10GHz, since the SW solutions have the same frequency for any $k_\parallel$ and any polarization when $\varepsilon = \mu = -1$ [17]. For a slab with a finite thickness, the coupling between two surfaces split the SW into two branches, as shown in Fig. 2(a). The upper (lower) branch of the SW spectrum is even (odd) with respect to the center $xy$ plane of the lens [18]. We note that the even-mode branch starts with a positive group velocity near the light line and then becomes negative group velocity after passing through a maximum with a zero group velocity. Figure 2(b) shows the density of states (DOS) of the SW excitations. In 2D systems, the SW has 1D dispersion properties and $DOS(\omega) \propto \partial k / \partial \omega = V_g^{-1}$. The correspondence between zero-group velocity in Fig. 2(a) and diverging DOS in Fig. 2(b) is obvious. In Fig. 2(b), we see a high DOS at 10 GHz, the frequency corresponding to the SW at a single interface, as expected. Another high DOS is due to the even-symmetry SW state with a zero-group velocity, which is a consequence of coupling between two interfaces, and is thus dependent on the slab thickness as Fig 2(b) shows. The "turning on" process of the monochromatic source will inevitably generate other frequency components. Given the divergence in the DOS, the maximum-frequency SW mode will be picked out together with the working frequency at 10 GHz, although its amplitude is much smaller than that of the working frequency wave because of the factor $(\omega - \omega_0 + i\eta)^{-1}$ [see Eq. (1)]. The beating of the input wave (denoted by $E_0 e^{i\omega_0 t}$) with this transient wave (denoted by $E_1 e^{i\omega_1 t}$) leads to a total field amplitude, $|E(t)| = \sqrt{|E_0|^2 + |E_1|^2 + 2\operatorname{Re}[E_1 E_0^* e^{i(\omega_1 - \omega_0)t}]}$, and this is the cause of the oscillation seen in Fig. 1. The oscillation frequency in Fig. 1 agrees *exactly* with $\bar{\omega} = \omega_1 - \omega_0$ found independently by the surface wave analysis. This beating frequency is higher in a thinner slab due to a larger splitting in the SW spectrum caused by a stronger interface coupling (see Fig. 2).

This particular transient wave can only be dissipated out by absorption. To illustrate this point, we plot the field amplitude and the time averaged energy current along the $x$ direction ($\bar{S}_x$) as the functions of $z$ (for $x = y = 0$) in Fig. 3(a) for this particular SW. Although the field is continuous across an air/lens interface, the energy current flows along opposite directions in different sides of the interfaces, since $\bar{S}_x = (k / 2\omega\mu) < \text{Re}(E_y^* E_y) >$ and $\mu$ has opposite signs in air and lens. In fact, at this particular frequency, $\bar{S}_x$ integrated in air regions completely cancel that in lens region, leading to a *net zero* (time averaged) energy current along *all* directions ($\bar{S}_z \equiv 0$ due to the evanescent wave nature). This is illustrated by the instantaneous pattern of the energy current distribution in $xz$-plane shown in Fig. 3(b). Dynamically, energy flows back and forth across the interfaces to form vortexes, which move along the $x$ direction at phase velocity $v = \omega / k_\parallel$, but no energy is transported along any direction. Since such a state does not transport energy, the corresponding fields of the state, once excited by the source, cannot be damped out through lateral transport. As a result, the oscillation will *never* stop in the case of $\gamma = 0$ [5]. With a finite $\gamma$, the lens will eventually absorb all the energy stored in such a state, and damp out the oscillations to give a final stable image [7]. This provides a complete picture to understand the image instability in super lens focusing [5-9].

We emphasize that such a mode is different from either a localized state or a standing wave, since our system is homogenous in the $xy$ plane. The phase velocity is zero in a standing wave, but is non-zero here. In fact, another state with an opposite phase velocity also possesses a *net zero* energy current. The negative refraction index of the lens, leading to a backward energy current, is the key to obtain such a state. We note the existence of such a state is inevitable (as long as $\varepsilon = \mu = -1$ at some frequency), independent on any specific form of $\varepsilon(f)$ and $\mu(f)$. Further works are necessary to identify other interesting implications of this unique mode.

We now study the 3D case. In 3D, the DOS for SW goes like $k\partial k/\partial\omega$. Since $k_{\parallel}$ extends to infinity at the frequency for focusing (10GHz here) but is finite at the upper band edge (see Fig. 2(a)), the strength of this vortex-state is relatively decreased compared with that of the focusing frequency, we thus expect weaker oscillations. Figure 4 shows the $|E_y^{3D}|$ calculated with the Green's function method at the exit surface of the lens and the image point, as the functions of time in unit of $t_0 = 2d/c$ (the traveling time in vacuum). Field values calculated by including only the propagating components are again shown for comparison. Figure 4 shows that the oscillation is indeed much weaker here although it is observable from a zoomed view. This is thus a marked dependence on dimensionality of the time evolution, a result probably not noted before.

In summary, we show that a vortex-like surface mode is responsible for the image oscillations observed previously for the 2D super lens focusing. The transient effect has a strong dependence on dimensionality. All results were obtained within a mathematically rigorous approach. We thank Jensen Li for helpful discussions. This work was supported by Hong Kong RGC through CA02/03.SC01 and National Basic Research Program of China (No. 2004CB719800).

**Figures**

**Fig. 1 (a) Image resolution** $w$ **as a function of time in 2D focusing, calculated with only propagating components (circles) and with all components (line). Here** $\gamma = 0.005$ **GHz,** $d = 10$ **mm. (b) Time dependent field amplitudes (in units of** $\mu_0 I_0$ **) for lenses with different parameters (** $\gamma$ **in GHz and** $d$ **in mm). Circles are calculated by including only propagating components for** $\gamma = 0.005$ **GHz,** $d = 10$ **mm.**

**Fig. 2 (a) TE mode SW spectra for lenses with thicknesses 10mm and 8mm. (b) Density of states of the surface wave excitation.**

**Fig. 3 (a) Distribution of** $|E_y|$ **(solid line, right scale) and** $\bar{S}_x$ **(dashed line, left scale) for the TE SW at the upper band edge frequency. (b) Instantaneous** $\vec{S}$ **distribution at the** $xz$-**plane for such an SW excitation [lens (air) region is denoted by gray (white) color].**

**Fig. 4 Time dependent field amplitudes (in units of** $\mu_0 P_0$ **) at the exit surface of the lens (dotted), at the image point (dashed), and at the image point but computed by including only propagating components (solid). Here** $\eta = 0.0005$ **GHz,** $\gamma = 0.002$ **GHz,** $d = 10$ **mm.**

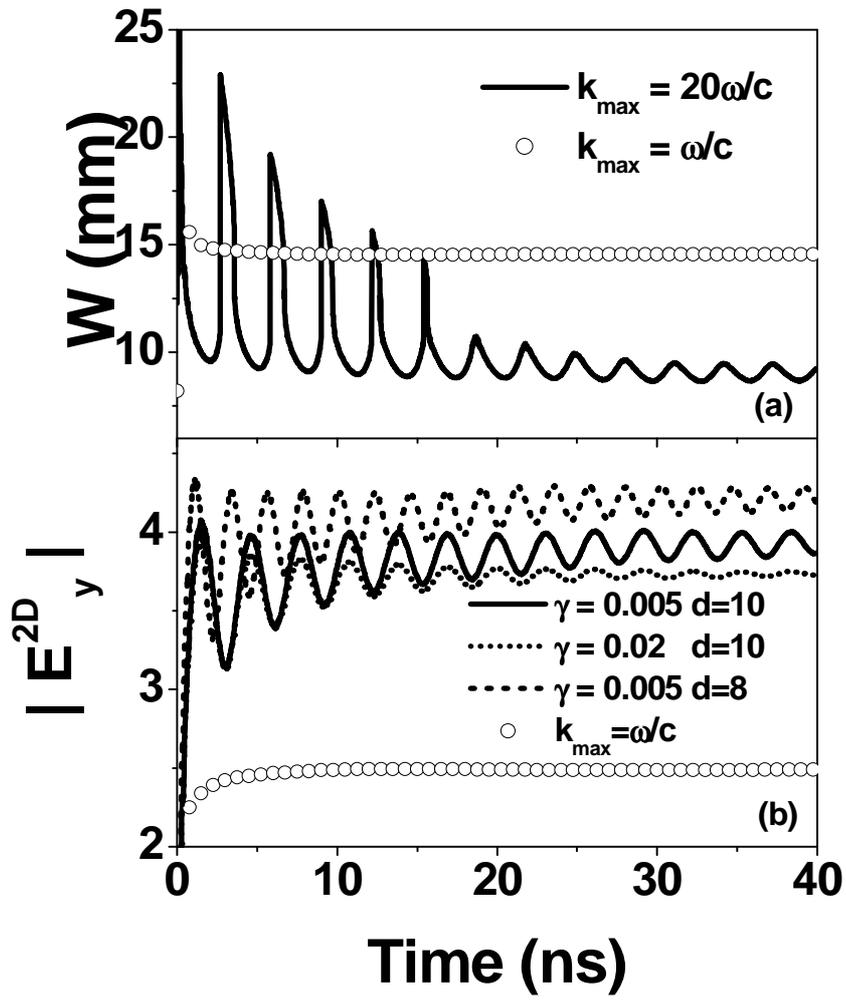

Fig. 1

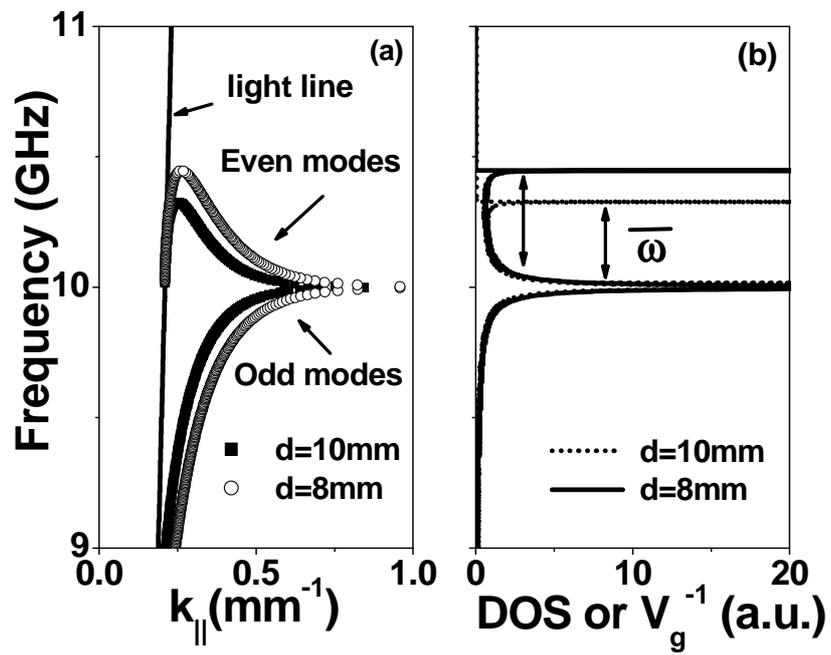

Fig. 2

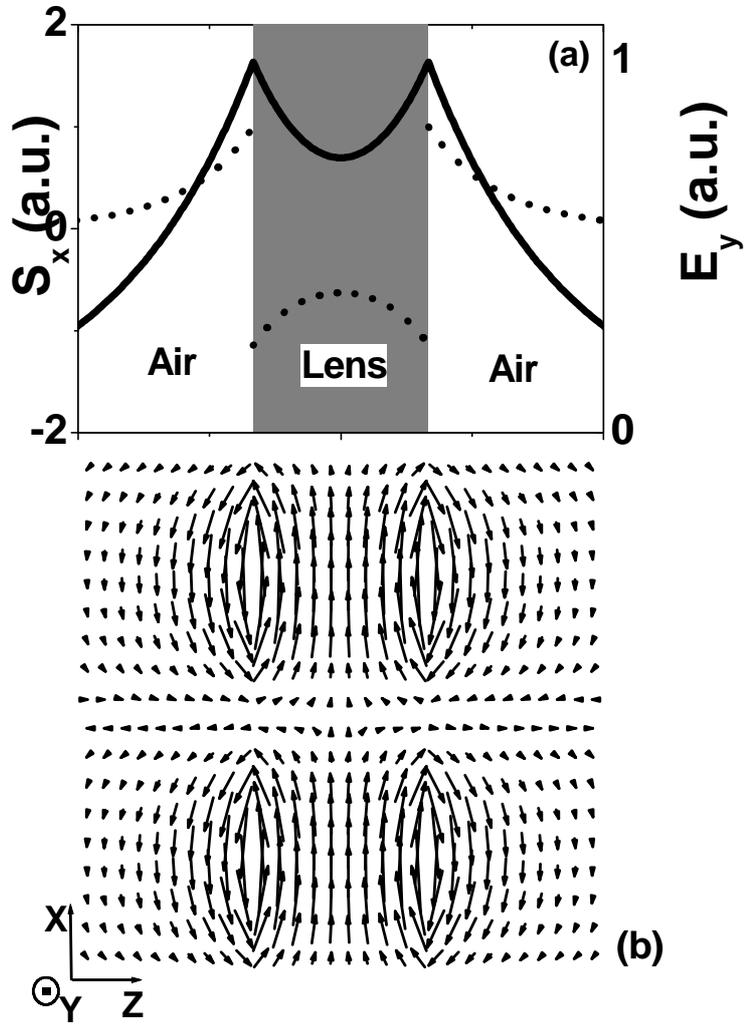

**Fig. 3**

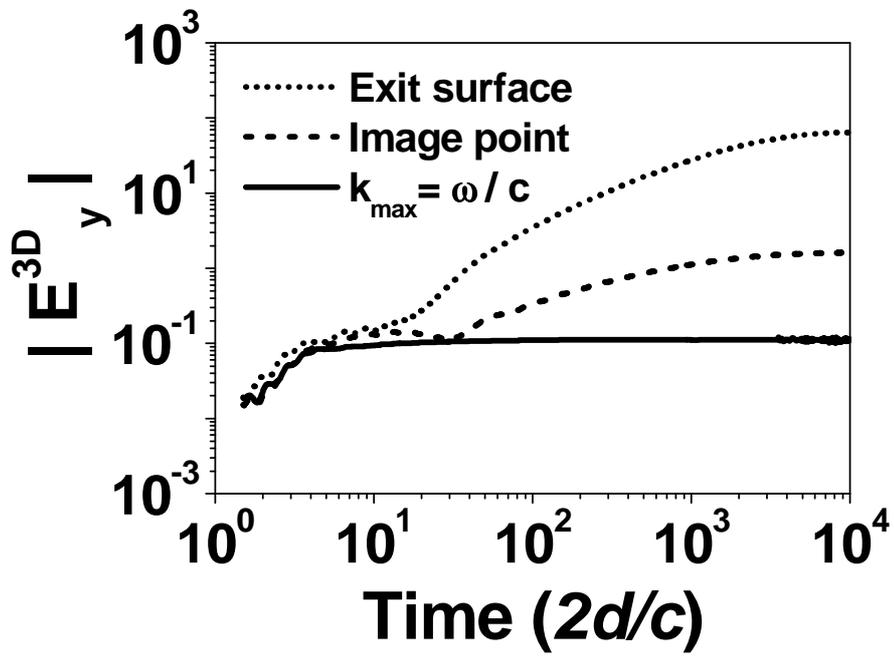

Fig. 4